
\magnification=1200\vsize=23.6 truecm\hsize= 15.4 truecm
\def\thf{\baselineskip=\normalbaselineskip\multiply\baselineskip
by 5\divide\baselineskip by 4}
\thf
\pageno=1

\centerline{\bf TRANSONIC ELASTIC MODEL FOR }
\centerline{\bf WIGGLY GOTO-NAMBU STRING}
\vskip 0.6 cm
\centerline{\bf Brandon Carter}
\vskip 0.4 cm
\centerline{D\'epartment d'Astrophysique Relativiste et de Cosmologie}
\centerline{C.N.R.S., Observatoire de Paris, 92 Meudon, France.}
\smallskip
\centerline{October, 1994.}
\vskip 0.6 cm
\parindent= 0 cm

{\bf Abstract.}  {\it The hitherto  controversial proposition that a
``wiggly" Goto-Nambu cosmic string can be effectively represented by an
elastic string model of exactly transonic type (with energy density
$U$ inversely proportional to its tension $T$)  is shown to have a firm
mathematical basis.}
\vskip 1 cm
\parindent=1 cm

For many years after work on cosmic strings was initiated by Kibble[1]
the subject was restricted to phenomena describable by a simple Goto-Nambu
model representing a vacuum defect of the non-conducting
kind whose prototype is the Nielsen-Olesen vortex[2]. More recently there
has been a new wave of cosmological interest in ``superconducting cosmic
strings", meaning phenomena due to a vacuum defect of the
current carrying kind whose prototype is the Witten vortex[3]. It
was suggested at an early stage[4] that a Witten vortex could be
represented  by an elegant current carrying model previously found by
Nielsen[5] as the outcome of a Kaluza Klein type projection from an ordinary
Goto-Nambu model in a background with an extra dimension.  However it soon
became clear that an adequate allowance even for weak currents
requires the use of elastic string models of a more general category[6] that
is characterised by {\it two} kinds of perturbation mode with generically
distinct propagation speeds[7] namely an extrinsic ``wiggle" (i.e.
worldsheet displacement) perturbation speed $c_{_{\rm E}}$ say and a sound
type ``woggle" (i.e. longitudinal) perturbation speed $c_{_{\rm L}}$ say (of
which the latter has no anologue in the internally strucureless Goto-Nambu
case for which the former is equal to the speed of light, set here to unity).

What disqualified the particular model obtained by the Nielsen mechanism[5]
as an accurate macroscopic representation of a Witten
vortex was the demonstration[8] that it is of {\it
permanently transonic} type, meaning that it has the
mathematically convenient but physically restrictive feature of equality of
the ``wiggle" and ``woggle" propagation speeds, i.e.
$$ c_{_{\rm E}}=c_{_{\rm L}}\ ,\eqno(1)$$
whereas Witten vortices were shown[9][10] to require representation by
elastic string models of {\it supersonic} type, as characterised by
the {\it strict} inequality  $c_{_{\rm E}}>c_{_{\rm L}}$.

A positive by-product of the analysis[8] leading to this negative conclusion
was the idea of an alternative application[11] for elastic string models of
the permanently transonic kind characterised by (1).
This was for the smoothed out macroscopic representation of an
``ordinary" (Goto-Nambu) cosmic string with a (thermal or other) spectrum of
``wiggles" that (for numerical or other reasons) one does not wish to
describe in detail. The aim of the present letter is to settle the
controversy that has arisen[12][13] about this proposal. Unlike the new
constructive proof presented here, the original argument[11] was merely
heuristic, while the confirmation[12] and its contradiction[13] did not deal
directly with the evolution of the smoothed out  worldsheet as is
done here, but merely with the question of whether
-- for an underlying Goto-Nambu model whose energy density $U$ and
tension $T$ have microscopically constant values $U=T=m^2$ say --
 the corresponding variably ``renormalised"[14][15] {\it average} values
 accurately satisfy the predicted[11] relation
$$UT=m^4 \ .\eqno(2)$$

The objection legitimately raised[13] against the previous
confirmation[12] was merely that it did not apply to quite the most general
type of ``wiggles", but only to a restricted (transversely generated) class.
However a more subtle issue in any such
verification is that the prediction[11] concerned the {\it effective}
energy density $U$ and tension $T$ of the smoothed out interpolating
string model, whereas the litigious calculations[12][13] were
concerned with  {\it averages} over
the ``wiggles" in the original Goto-Nambu model. The delicate question
is which of variously weighted averages $\overline\epsilon$,
$\langle\varepsilon\rangle$, ... of the microscopic energy density
$\varepsilon$ is to be identified with the ``renormalised" effective
value, $U$, and likewise for the tension. Since getting
the ``correct" answer depends on using the ``correct" definition,
the issue of how accurately (2) holds is to some extent semantic.

The new (constructive) derivation of the smoothed out transonic elastic
string representation does not involve any such averaging procedure and thus
bypasses the semantic issue implicit in the preceding debate[12][13]. The
apparent discrepancy has moreover been recently resolved by Martin[16], who
has shown that the predicted relation (2) is in fact obtained (even for non
transversely generated wiggles) provided one uses the appropriate ``effective"
average as defined for energy by $\overline\varepsilon=E/x$ where
$E=\int\varepsilon d\sigma$ is the total energy integrated along the length
$\sigma=\int d\sigma$ of a ``wiggly" segment of the worldsheet, and $x$ is
the shortcut (background coordinate) distance between its ends, which is
given, in terms of the cosine $x^\prime=dx/d\sigma$ of the angle between the
``wiggly" path and the shortcut interpolation, by $x=\int x^\prime d\sigma$.
In terms of  the ``path" averages defined by $\langle\varepsilon\rangle=
\sigma^{-1} \int\varepsilon d\sigma $ and $\langle x^\prime\rangle=
\sigma^{-1} \int x^\prime d\sigma$ the required ``effective" average is
$\overline\varepsilon=\langle\varepsilon\rangle / \langle x^\prime \rangle$.
The reported deviations[13] are attributable to the use of a slightly
different ``intermediate" average defined by $\widetilde\varepsilon=
\langle\varepsilon/x^\prime\rangle$.

Showing how to fix the definitions so as to get a formula of the predicted
form (2) is by itself not quite sufficient to show that the proposed
model[11] can actually achieve its purpose of smoothly interpolating
through the ``wiggles" of the underlying Goto-Nambu string. To do
this, one must
go back to the equations of motion which, both for a Goto-Nambu string and for
strings of the prefectly elastic category[6] needed here (though not for
more general multiply conducting models[17]) are expressible completely (so
long as there are no self intersections) just by the standard force balance
relation[8]
$$\overline\nabla_{\!\mu}\overline T{^{\mu\nu}}=\overline f{^\nu}\ ,\eqno(3)$$
where $\overline\nabla_{\!\mu}$ denotes the worldsheet projected covariant
differentiation operator whose  definition is recapitulated below,
while $\overline f{^\mu}$ is the external electromagnetic or other
(e.g. material drag[18]) force density, which  will be assumed  here
to vanish, and $\overline T{^{\mu\nu}}$ is the surface
stress momentum energy density of the string. The latter, which is
tensorial with respect to the spacetime coordinates $x^\mu$ ($\mu$=0,1,2,3,4),
is given in terms of its internal analogue $t^{ab}$ which is tensorial with
respect to internal worldsheet coordinates $\sigma^a$ (a=0,1) by
$$\overline T{^{\mu\nu}}=t^{ab}x^\mu_{\, ,a} x^\nu_{\, ,b}\hskip 1 cm
t^{ab}=2{\partial{\cal L}\over\partial h^{ab}}+{\cal L} h^{ab} \ ,
\eqno(4)$$
(using a comma for partial diferentiation) where ${\cal L}$ is the surface
Lagrangian density (which is just a constant in the Goto-Nambu case) and
$h^{ab}$ is the contravariant version of the induced metric, whose spacetime
projection gives the fundamental tangential projection tensor
$$\eta^{\mu\nu}= h^{ab}x^\mu_{\, ,a} x^\nu_{\, ,b}\ , \hskip 1 cm
h_{ab}= g_{\mu\nu}x^\mu_{\, ,a} x^\nu_{\, ,b} \ ,\eqno(5)$$
that is needed for defining the tangentially projected derivation operator
in (3) and also the orthogonal projection tensor $\perp^{\!\mu}_{\,\nu}$
that will be needed below:
$$\overline\nabla_{\!\mu}=\eta^\nu_{\ \mu}\nabla_{\!\nu}\ ,\hskip 1 cm
\perp^{\!\mu}_{\,\nu}=g^\mu_{\ \nu}-\eta^\mu_{\ \nu}\ .\eqno(6)$$
The corresponding Dirac distributional stress momentum energy density
tensor $\hat T{^{\mu\nu}}$ say over the background is given by
$$ \hat T^{\mu\nu}=\Vert g \Vert^{-1/2}\int\!  \overline T{^{\mu\nu}}
\delta^{\rm 4}[x-x\{\sigma\}]\,\sqrt{\Vert h \Vert}\,
 d^{\rm 2}\!\sigma \ . \eqno(7)$$
It is easily verified that vanishing of the force term in the regular
differential equation (3) is equivalent to conservation (in the usual sense)
of this distribution:
$$\overline f{^\mu}= 0 \hskip 0.6 cm  \Leftrightarrow
\hskip 0.6 cm \nabla_{\!\mu}\hat T{^{\mu\nu}}=0\ . \eqno(8)$$

In a generic state, $\overline T{^{\mu\nu}}$ will have
timelike and spacelike unit
worldsheet tangential eigenvectors $u^\mu$ and $v^\mu$ say that define a
preferred orthonormal tangent frame in terms of which one obtains
$$\overline T{^{\mu\nu}}= U u^\mu u^\nu- T v^\mu v^\nu\ ,\hskip 1 cm
\eta^{\mu\nu}=-u^\mu u^\nu+v^\mu v^\nu \ ,\eqno(9)$$
in which the eigenvalues are the energy density $U$ and tension $T$ referred
to above. The permanently transonic model proposed for representing the
smoothed interpolating worldsheet is obtainable, in acordance with (4) from
a Lagrangian density of the form ${\cal L}=-m^2\sqrt{1+ h^{ab}\psi_{,a}
\psi_{,b}\,}\,$ where $\psi$ is an independently variable stream function
(whose gradient gives the direction of the spacelike eigenvector $v^\mu$)
which leads to $U=-{\cal L}$ and $T=-m^4/{\cal L}$, in evident accord with
(2). The underlying Goto Nambu model that one wants to represent this way is
given more simply by ${\cal L}=-m^2$ which gives $U=T=m^2$ while leaving the
eigendirections unspecified. In both the Goto Nambu and the generic elastic
cases, (9) can be conveniently rewritten in the form
$$\overline T{^{\mu\nu}}={_1\over^2}\big(\beta_{_+}^{\,\mu}\beta_{_-}^{\,\nu}
+ \beta_{_-}^{\,\mu}\beta_{_+}^{\,\nu}\big)\ , \hskip 1 cm
\beta_{_\pm}^{\,\mu}=\sqrt{U}u^\mu \pm \sqrt{T} v^\mu \ .\eqno(10)$$
Taking the orthogonal projection (by contraction with $\perp^{\!\mu}_{\,\nu}
 $) of the dynamical relations (3) and combining the result with the kinematic
identity
$$\perp^{\!\mu}_{\,\nu}\big(\beta_{_+}^{\,\rho}\nabla_{\!\rho}
\beta_{_-}^{\,\nu} - \beta_{_-}^{\,\rho}\nabla_{\!\rho}\beta_{_+}^{\,\nu}
\big)=0 \ ,\eqno(11) $$
(which  holds just as Frobenius type integrability condition for the
vectors $\beta_{_\pm}^{\,\nu}$ to be and to remain worldsheet tangential)
one obtains the extrinsic part of the dynamical equations (3) in the neat
characteristic form
$$\perp^{\!\mu}_{\,\nu}\beta_{_\pm}^{\,\rho}\nabla_{\!\rho}
\beta_{_\mp}^{\,\nu}= \perp^{\!\mu}_{\,\nu}\overline f{^\nu}\ ,\eqno(12)$$
which is valid for any kind of elastic string model, and from which it can
be seen that  $\beta_{_\mp}^{\,\nu}$ are actually bicharacteristic and hence
by their construction (10) that relative to the preferred frame specified by
$u^\mu$ the propagation speed for extrinsic ``wiggle" perturbations will be
given by the generally valid formula $c_{_{\rm E}}= \sqrt{T/U}$.
When the force density $\overline f{^\nu}$ is set to zero, it can be seen
that (12) is expressible in terms of arbitrarily rescaled bicharacteristic
vectors $\ell_{_\mp}^{\,\mu}=\alpha\beta_{_\mp}^{\,\mu}$ simply as
$$\perp^{\!\mu}_{\,\nu}\ell_{_\pm}^{\,\rho}\nabla_{\!\rho}
\ell_{_\mp}^{\,\nu}= 0 \ .\eqno(13)$$

In the familiar Goto-Nambu case for which the bicharacteristic vectors are
null there remains no tangentially contracted part of the equations of
motion, which are completely contained in (12). Thus  there
is an indeterminacy which can be resolved,
taking advantage of the rescaling freedom in (13), by taking the
bicharacteristic vectors to be Lie transported along each other. The
orthogonal projection in (12) can thereby be removed, leaving the equations
of motion in the fully determinate form
$$\ell_{_\pm}^{\,\rho}\nabla_{\!\rho}\ell_{_\mp}^{\,\nu}=0\ .\eqno(14)$$
This system is well known to be completely soluble in a flat background in
terms of characteristic coordinates $\sigma^{_+}$ and $\sigma^{_-}$ such
that
$$\ell_{_\pm}^{\,\rho}={\partial x^\mu\over\partial\sigma^{_\pm} }
\ ,\eqno(15)$$
by the familiar ansatz[19] (on which the calculations referred to
above[12][13] were based) given, using Minkowski coordinates $x^\mu$, by
$$x^\mu\{\sigma^{_+},\sigma^{_-}\}=
x_{_+}^{\,\mu}\{\sigma^{_+}\}+x_{_-}^{\,\mu}\{\sigma^{_-}\}\eqno(16)$$
for a pair of generating curves $x_{_\pm}^{\,\mu}\{\tau\}$ that can
``wiggle" freely subject to the restriction that their tangent vectors $\dot
x_{_\pm}^{\,\mu}=d x_{_\pm}^{\,\mu}/d\tau $ remain everywhere  null, i.e.
$\dot x_{_+}^{\,\mu} \dot x_{_+\mu}=0$ and
$\dot x_{_-}^{\,\mu} \dot x_{_-\mu}=0$. Thus if the
parameter $\tau$ is chosen to be just the Minkowski
time coordinate $x^{_0}$, so that $\dot x_{_\pm}^{\,_0}=1$, the space
vectors with components $\dot x_{_\pm}^{\,i}$ for $i=1,2,3$ will
lie on the surface of the (Kibble Turok) unit sphere.

To deal with the generic elastic case in which $T$ is less that $U$ so
that the bicharacteristic vectors $\beta_{_\pm}^{\,\rho}$ will be not null
but timelike, it is convenient to  use a rescaling given by
 $$\beta_{_\pm}^{\,\rho}=\sqrt{U-T}\, \ell_{_\pm}^{\,\rho}\ ,\eqno(17)$$
so as to obtain a pair of bicharacteristic vectors  $\ell_{_\pm}^{\,\rho}$
that have unit normalisation. In this generic case the internal dynamical
equations obtained by tangential projection of (3) are no longer trivial,
but, in the special case (2) considered here, the pair of internal
equations got by contracting (3) with the independent bicharacteristic
vectors can be cast into the particularly simple form
$$\overline\nabla_{\!\mu}\Big( (U-T) \ell_{_\mp}^{\,\mu}\Big)
=-\ell_{_\pm}^{\,\nu} \overline f_\nu\ . \eqno(18)$$
By a more elaborate manipulation, these simple surface divergence equations
replaced by what for the present purpose is the more useful
combined equation
$$(U-T)\eta^{\mu}_{\ \nu}\ell_{_\pm}^{\,\rho}\nabla_{\!\rho}
\ell_{_\mp}^{\,\nu}= \big(\eta^{\mu\nu}+\ell_{_\mp}^{\,\mu}
\ell_{_\mp}^{\,\nu}\big)\overline f{_\nu}\ .\eqno(19)$$
By showing that the tangent vectors $\ell_{_\pm}^{\,\mu}$ are
bicharacteristic for internal``woggles" not just for extrinsic ``wiggles",
this demonstrates the transonicity property (1).

It is obvious by (6) that, when the force term on the right is absent,
recombining the intrinsic dynamical equation (19)
with its extrinsic analogue (13) reconstitutes the complete (unprojected)
set of dynamical equations in {\it the same} simple form (14) as was obtained
by judicious use of the gauge freedom for the null characteristic vectors
of the Goto-Nambu case. This crucial result implies, as before, that in
a Minkowski background the general solution will be obtainable in terms of
characteristic coordinates  $\sigma^{_+}$ and $\sigma^{_-}$
(corresponding to preferred internal time and space coordinates
$\tau=\sigma^{_+}+\sigma^{_-}\ $, $\ \sigma=\sigma^{_+}-\sigma^{_-}\ $)
by an ansatz of the same form (16) as in the Goto-Nambu case, the only
difference being that, instead of being null, the separate generating curves
are now required to be {\it timelike}. In order for the characteristic
vectors given by (15) as $\ell_{_\pm}^{\,\mu}=\dot x_{_\pm}^{\,\mu} $ to
satisfy the unit normalisation condition, the separate generating curves
would also need to be restricted to have a {\it proper time} parametrisation,
but it is evident that this is not obligatory in order for (16) to be
a solution. An alternative is the {\it standard} parametrisation
given by the background time coordinate $x^{_0}$  on each
generating curve, which  (with $\tau=x^{_0}$) gives $\dot x_{_\pm}^{\,_0}=1$:
for the generators to be timelike,  the 3-vectors  $\dot x_{_\pm}^{\,i}$
must then lie, not on the surface as in the Goto-Nambu case, but
{\it inside} the (Kibble Turok)  unit sphere.

To see how the solution that has just been given for the worldsheet of the
transonic elastic model  can be used to provide a smoothed
interpolation through the ``wiggles" of the worldsheet of an underlying
Goto-Nambu model, it now suffices to use an idea introduced by Smith and
Vilenkin[20] for the purpose of numerical computation, for which one needs
to replace the exact continouous description of the worldsheet by a
discrete representation. The Smith Vilenkin method is simply to use a
pair of discrete sets of sampling points  $x_{_\pm\rm r}^{\,\mu}=
 x_{_\pm}^{\mu}\{\sigma_{\rm r}\}$ determined by a corresponding
discrete set of parameter values $\sigma_{\rm r}$ on
the generating curves of the exact representation (16).
This provides a ``diamond lattice" of sample points given (for integral
values of r and s) by
$$x^\mu_{\rm rs}={_1\over^2}\big( x_{_+\rm r}^{\,\mu}
+x_{_-\rm s}^{\,\mu}\big)     \ ,\eqno(20) $$
that will automatically lie {\it exactly} on the ``wiggly"  Goto-Nambu
worldsheet (16), which is thus represented to any desired accuracy by
choosing a sufficiently dense set of sampling parameter values $\sigma_{\rm
r}$ on the  separate ``wiggly" null generating curves $x_{_\pm}^{\,\mu}
\{\sigma\}$. To construct a corresponding smoothed out worldsheet
it suffices to consider the chosen set of sample points
$x_{_\pm\rm r} ^{\,\mu}=x_{_\pm}^{\mu}\{\sigma_{\rm r}\}$
on the separate ``wiggly" null generators  to be
sample points on a pair of {\it smoothed out} -- and thus no longer null but
timelike -- {\it interpolating curves}. Using these
{\it timelike} interpolating curves as generators,
the {\it same} ansatz (16) can now be used to construct another
smoother {\it interpolating worldsheet} that will satisfy (14) and
therefore will be {\it an exact solution} of the dynamic
equations for the transonic elastic string model.
(The form (16) of the general solution also shows incidentally that a
similarly smoothed out interpolation  for the worldsheet of a
 ``wiggly" {\it and ``woggly"}  elastic string of transonic type will
be provided by an effective model of {\it the same} transonic type.)

The (less ``wiggly") elastic string worldsheet given by this construction
will obviously be an even better approximation to the (more ``wiggly")
Goto-Nambu worldsheet than the original Smith Vilenkin lattice
representation, which itself could already be made as accurate as desired
by choosing a sufficiently high sampling resolution. No matter how far it
is extrapolated to the future, the smoothed elastic string worldsheet can
never stray significantly from the underlying ``wiggly" Goto-Nambu worldsheet
it is designed to represent (at least in the absence of background curvature
whose effect remains a topic for future investigation) because the exact
worldsheet and the elastic interpolation will always coincide precisely at
each point of their shared  Smith Vilenkin lattice (20).  This highly
satisfactory feature of guaranteed error cancellation in the long run could
not be improved, but would only be spoiled, by any ``deviation" from (2).

The arbitrarily accurate worldsheet matching property that has just been
demonstrated, does not depend on the mass scale $m$ in the action for the
transonic elastic model being the same as that of the underlying Goto-Nambu
model, because for both kinds of model the actual equations of motion are
{\it scale independent}. The same mass scale is however needed if one wants
agreement of the effective energy density and tension of the smoothed elastic
model with corresponding (suitably defined) averages over the ``wiggles"
in the underlying Goto-Nambu model.

Having thus provided a solid (directly constructive) basis for the claim
that the permanently transonic elastic string model  provides an
excellent description of the effect of microscopic wiggles in a
Goto-Nambu string so long as self intersections remain unimportant
(as was assumed throughout the debate[12][13] discussed above), it must
be emphasised that the neglect of such intersections will not be
justified when the effective temperature[11][17] of the ``wiggles" is too
high (as will presumably be the case[21] during a transient
period immediately following the string - forming phase transition).
Such intersections will produce microscopic loops, of which some will
escape. Estimation of the dissipative force density  $\overline f{^\nu}$
that would be needed to allow for the ``cooling" effect of such
losses remains a problem for future work.

\bigskip\parindent=0 cm
The author wishes to thank A. Vilenkin and X. Martin
for helpful conversations.
\medskip

\bigskip\parindent = 0 cm
 {\bf References.}
\medskip

[1] T.W.B. Kibble, {\it J.Phys.} {\bf A9}, 1387 (1976).
\smallskip
[2] H.B. Nielsen, P. Olesen, {\it Nucl. Phys.} {\bf B61}, 45 (1973).
\smallskip
[3] E. Witten, {\it Nucl. Phys.}{\bf B249}, 557 (1985).
\smallskip
[4] N.K. Nielsen, P. Olesen, {\it Nucl. Phys.} {\bf B291},
829 (1987).
\smallskip
[5] N.K. Nielsen, {\it Nucl. Phys.} {\bf B167} 248 (1980).
\smallskip
[6] B. Carter, {\it Phys. Lett.} {\bf B224 }, 61 (1989).
\smallskip
[7] B. Carter, {\it Phys. Lett.} {\bf B228}, 446 (1989).
\smallskip
[8]  B. Carter,   in {\it Formation and Evolution of Cosmic Strings,}
ed. G. Gibbons, S. Hawking, T. Vachaspati, pp143-178 (Cambridge U.P., 1990).
\smallskip
[9] P. Peter, {\it Phys. Rev.} {\bf D45}, 1091 (1992).
\smallskip
[10] P. Peter, {\it Phys. Rev.} {\bf D47}, 3169 (1993).
\smallskip
[11] B. Carter, {\it Phys. Rev.} {\bf D41}, 3869 (1990).
\smallskip
[12] A. Vilenkin, {\it Phys. Rev.} {\bf D41}, 3038 (1990).
\smallskip
[13] J.Hong, J.Kim, P. Sikivie, {\it Phys. Rev. Lett.} {\bf 69},
2611 (1980).
\smallskip
[14] B. Allen, E.P.S. Shellard, {\it Phys. Rev. Lett.}
{\bf 64}, 119 (1990).
\smallskip
[15]  E.P.S. Shellard, B. Allen, in {\it Formation and Evolution of Cosmic
Strings,} ed. G. Gibbons, S. Hawking, T. Vachaspati, pp421-448  (Cambridge
U.P., 1990).
\smallskip
[16] X. Martin, preprint (Observatoire de Paris, Meudon, 1994).
\smallskip
[17] B. Carter, {\it Nucl. Phys.} {\bf B412}, 345 (1994).
\smallskip
[18] A. Vilenkin, {\it Phys. Rev.} {\bf D43}, 1060 (1991).
\smallskip
[19] T.W.B. Kibble, N. Turok, {\it Phys. Lett} {\bf B116}
141 (1982).
\smallskip
[20] A.G. Smith, A. Vilenkin, {\it Phys. Rev.} {\bf D36} 990
(1987).
\smallskip
[21] B. Carter, M. Sakellariadou, X. Martin, {\it Phys. Rev.} {\bf D50},
628 (1994).

\end